\documentclass{kapproc} % Computer Modern font calls

\usepackage{procps} 

\usepackage[dvips]{graphicx}

\upperandlowercase

\kluwerbib % will produce this kind of bibliography entry:

\begin{document}

\articletitle[Secular evolution versus hierarchical merging]
{Secular evolution versus hierarchical merging: 
galaxy evolution along the Hubble sequence, in the field and rich environments}

\author{Francoise Combes}
 
\affil{LERMA, Observatoire de Paris, 61 Av. de l'Observatoire,
F-75014, Paris, France}

\begin{abstract}
In the current galaxy formation scenarios, two physical 
phenomena are invoked to build disk galaxies: hierarchical mergers
and more quiescent external gas accretion, coming from
intergalactic filaments. Although both are
thought to play a role, their relative importance is not known
precisely. Here we consider the constraints on these scenarios
brought by the observation-deduced 
star formation history on the one hand,
and observed dynamics of galaxies on the other hand: 
the high frequency of bars and spirals, the high 
frequency of perturbations such as lopsidedness, warps, or polar rings.
  All these observations are not easily
reproduced in simulations without important gas accretion.
N-body simulations taking into account the mass exchange between 
stars and gas through star formation and feedback, can reproduce the data, 
only if galaxies double their mass in about 10 Gyr through gas accretion.
Warped and polar ring systems are good tracers of this accretion,
which occurs from cold gas which has not been virialised in
the system's potential. The relative importance of these phenomena 
are compared between the field and rich clusters.
The respective role of mergers and gas
accretion vary considerably with environment.
\end{abstract}


\section{Introduction}

Galaxies grow from small mass systems, the first structures to be
unstable after recombination are of the order of a globular cluster mass
($\sim$ 10$^6$ M$_\odot$). Then, according to the hierarchical scenario
of galaxy formation, small systems merge to form larger and larger systems,
and in the same time, small structures accrete gas mass and dark
matter from larger
gaseous structures in the shape of filaments in the cosmic web.

The relative importance of these two essential ways to build
galaxies, through mergers or external gas accretion, is still 
unprecised in the numerical simulations, since it depends on 
many unknown parameters of the baryonic physics, gas dissipation,,
star formation efficiency, feedback, gas re-heating, etc..
However the dynamical history of galaxies strongly depends
on these processes, and it might be possible to find constraints
on them from the observations, locally and at various redshifts,
of the dynamical states of galaxies.

Interaction with massive companions and mergers tend to
heat the stellar component of galaxies, and to form the spheroids,
either increase the mass of the central bulge, or even
transform the system in a giant ellipticals in case of
major mergers. On the contrary, gas accretion can replenish
the young disk in spiral galaxies, and rejuvenate spiral
or bar waves, and reduce the bulge-to-disk ratio. 
The thickness of galaxy disks, and their ability to
maintain spiral structure and asymmetries is a 
tracer of their history, in terms of interactions
or gas accretion. I will review here the recent progress in 
obtaining these constraints, both on the observational
side, which provides statistics of the dynamical state
of spiral galaxies, and on the theoretical side, which
provides the interpretation through predictions of this
dynamical state under several hypothesis about the
past history of galaxies.

Section 2 reviews the star formation history of
galaxies in the field, and Section 3 the constraints
from the dynamics: bars, warps, polar rings and asymmetries.
Section 4 considers the same phenomena in clusters,
and conclusions about secular evolution as a function of
environment are drawn in Section 5.

\section{Star formation history}

Models of the chemical evolution of the Milky Way suggest that the observed 
abundances of metals require a continuous infall of gas with metallicity  
about 0.1 times the solar value. This can solve the  well-known G-dwarf 
problem, i.e. the observational fact that the metallicities of most 
long-lived stars near the Sun lie in a relatively narrow range 
(-0.6 $<$ [Fe/H] $<$ +0.2, Rocha-Pinto \& Maciel 1996). 
The infall of gas is also supported by the constant or
increasing star formation rate (SFR) scenario inferred from
the local distribution of stars (e.g. Haywood et al. 1997). Other abundance
 problems require also an infall rate integrated over the entire disk of 
the Milky Way of a few solar mass per year at least (Casuso \& Beckman 2001).
 This infall dilutes the enrichment arising from the production of heavy
 elements in stars, and thereby prevents the metallicity of the 
interstellar medium from increasing steadily with time. Some of this gas
 could come from the High Velocity Clouds (HVC) infalling onto our galaxy 
disk (Wakker et al 1999).

The High-Velocity Clouds (HVCs) observed in the Galactic neighbourhood, at 
least those not included into the Magellanic Stream more akin to tidal debris,
 have been proposed to be remnants of the formation of the galaxies in the
 Local Group (Blitz et al 1999). With distances much larger than previously
 assumed (i.e. 1 Mpc instead of 100 kpc), their gas masses  could represent
 a large fraction of the total baryonic mass of the Local Group. This hypothesis
 is supported by observational evidence that their kinematical centre is the
 Local Group barycentre (Blitz et al 1999). Within this hypothesis, HVCs
 can well explain the evolution of the light elements in the Galaxy, and the
 G-dwarf problem (Lopez-Corredoira et al 1999). The present infall rate of
 gas is estimated to be 7.5 Mo/yr (Blitz et al 1999). This is the right order 
of magnitude to double the baryonic mass of the Galaxy in 10 Gyr time-scale.
 The HVCs, whose properties are very similar to the higher redshift Ly$\alpha$
 forest clouds, may thus form a significant constituent of baryonic, and of 
non-baryonic, dark matter.

It is now possible to investigate the star formation evolution in nearby 
galaxies as well: 
Worthey \& Espana (2004) with HST colour-magnitude diagrams derived a stellar 
abundance distribution for M31. The results are quite  similar to what is 
observed in the solar neighbourhood and they concluded that closed-box models 
in M31 suffer a G-dwarf problem even more severe than in the Milky Way.

The star formation rate in the Milky Way disk has remained of the same order 
of magnitude over the galaxy life-time (Rana \& Wilkinson 1986), although some
 temporal fluctuations can be identified (Rocha-Pinto et al 2000). This 
appears to be a general results for spiral galaxies in the middle of the 
Hubble sequence (e.g. Kennicutt 1983, Kennicutt et al 1994).

\begin{figure}[ht]
\centerline{\includegraphics[width=10cm]{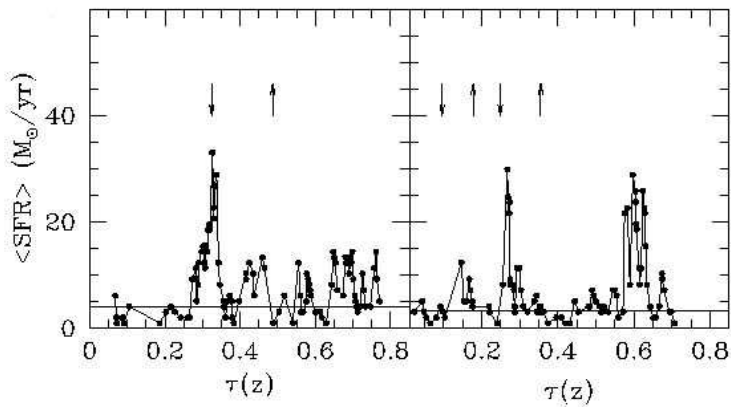}}
\caption{Star formation rate history of two
 galactic-like objects modelled by Tissera (2000).
The arrival of a companion is indicated 
an arrow pointing up, and the merging
of the baryonic cores by
an arrow pointing down.}
\label{tissf2}
\end{figure}

\begin{figure}[ht]
\centerline{\includegraphics[width=10cm]{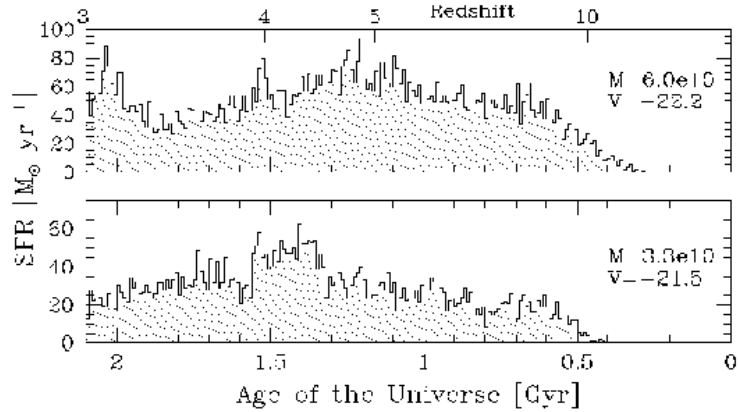}}
\caption{Star formation histories of Lyman-break galaxy models
at $z=3$, by Nagamine et al (2004). The mass of the galaxy in
$h^{-1}$ M$_\odot$ and its rest-frame V-band magnitude are indicated
at right.}
\label{naga-f8}
\end{figure}

To maintain the same order of magnitude for star formation rate requires 
external gas accretion, since an isolated galaxy must have an exponentially
 decreasing star formation rate, even taking into account stellar mass loss. 
Numerical simulations of galaxy evolution, in a 
general cosmological frame, have shown that indeed a typical spiral galaxy 
in average environment, maintains a constant level of star formation rate,
 with superposed bursts (Tissera 2000, Nagamine et al 2004). 
While simulations of isolated galaxy
 mergers reveal how star formation is triggered by interaction and mergers 
(e.g. Mihos \& Hernquist 1996), they cannot maintain a constant average level
 of star formation, as in cosmological simulations.

Figures \ref{tissf2} and \ref{naga-f8} shows several histories of star 
formation for galaxy-like objects at $z=0$, from Tissera (2000) and at $z=3$, 
Nagamine et al. (2004). The history has been obtained by
 reconstructing the merging tree, in identifying the progenitors of the present
 galaxy with the particles it is made of in the simulation, a companion being 
defined by having at least 10\% of the main galaxy mass (the accretion of
 baryonic clumps are assimilated to external gas accretion). Each time such a 
companion enters the main halo, an up arrow is drawn, and each merger is traced 
by a down arrow. Several phenomena can be emphasized: a large fraction of gas is
 accreted between mergers, and this contribute an essential part of the star 
formation. The tidal interaction triggers bursts of star formation, which in 
general are delayed with respect to the first passage, but the bursts are not 
only fuelled by the accreted companion gas, but mainly from the main galaxy gas
 driven in by the tidal forces. In that sense, the external gas accretion in 
between mergers also fuels the merger-triggered starbursts.

Observations of stellar populations in a large sample of local galaxies (SDSS,
 Heavens et al 2004, Jimenez et al 2004) have shown that massive galaxies have 
formed most of  their stars at early times, while dwarf galaxies are forming 
their stars efficiently only now. Only intermediate mass galaxies could have in
 average maintained their star formation rate over a Hubble time. The high 
efficiency of star formation in massive galaxies underlines the role of environment,
 and the availability of  high gas accretion. The rate of gas consumption is then 
high, and the galaxies run out of gas quickly. On the contrary, far from deep 
potential wells, dwarf galaxies need billion years to reach the threshold of star
 formation, and convert their gas only now. 

\begin{figure}[ht]
\centerline{\includegraphics[width=10cm]{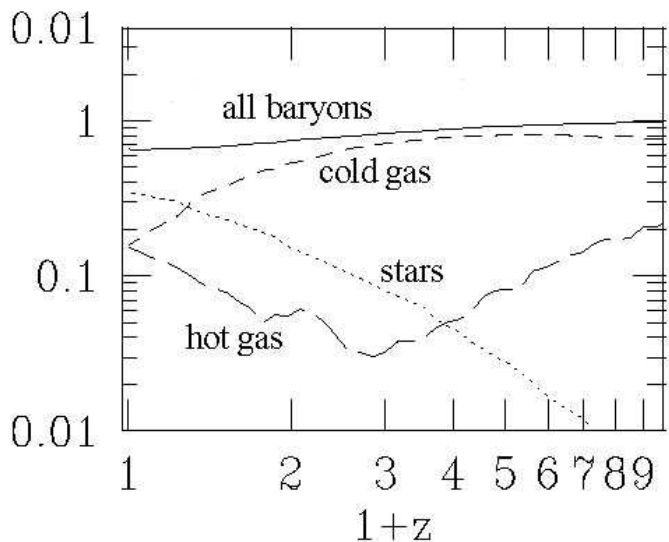}}
\caption{The fraction of cold and hot gas, and the history
of star formation in halos of size $V_c=$ 220 km/s at $z=0$, in 
a standard CDM model, computed with semi-analytic models
by Somerville \& Primack (1999). }
\label{sp99-f5}
\end{figure}

In semi-analytic models, an essential free parameter to reconstruct merger trees 
and mass assembly in galaxies, in the fraction of accretion, and this can change 
considerably the final angular momentum of the system (Wechsler et al 2002).
Figure \ref{sp99-f5} illustrates the time evolution of the baryonic content
of halos that correspond to a "local group" ($V = 220$ km/s) sized halo at $z=0$.
This evolution corresponds to the best fit of the observations, i.e. 
the luminosity function of galaxies and the Tully-Fisher relation.
The star formation rate has been calibrated, so that the star formation
efficiency is constant with redshift (Somerville \& Primack 1999).
Note that the fraction of available gas in the form of
cold phase remains quite high until $z=0$.

\begin{figure}[ht]
\centerline{\includegraphics[width=10cm]{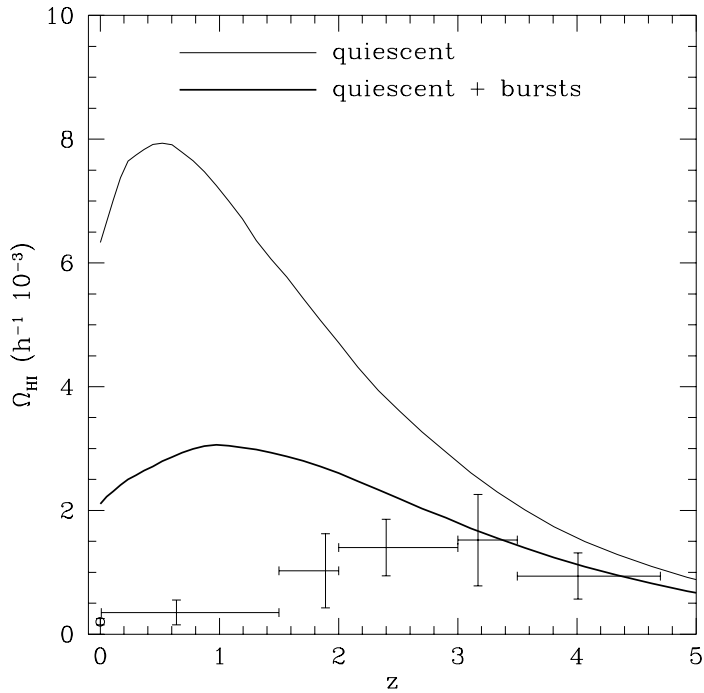}}
\caption{Cold gas density versus redshift, computed in
semi-analytic models by  Somerville \& Primack (1998). The bold
solid line includes starbursts, while the light line only
quiescent star formation.
Data points show the density in the form of HI estimated
from observations of DLA (Storrie-Lombardi et al. 1996). The point at
$z=0$ is from local HI observations (Zwaan et al 1997). }
\label{sp98-f6}
\end{figure}

As far as the global gas content is concerned, constraints can
be found in the frequency and column densities of the damped
Lyman-$\alpha$ systems (DLA). Already at high redshift, the models
are limited by the observations, which then constrain the
star formation rate (fig \ref{sp98-f6}). A strong starburst
mode is required in addition to quiescent star formation mode, 
as triggered by the frequent mergers around $z \sim 2-1$. 
Note that the observational points only correspond to HI gas
of column density larger than 2 10$^{20}$ cm$^{-2}$, ignoring
molecular or ionised gas, that could bring the model
closer to observations at low redshift.

\section{Constraints from the dynamics}

\begin{figure}[ht]
\centerline{\includegraphics[width=10cm]{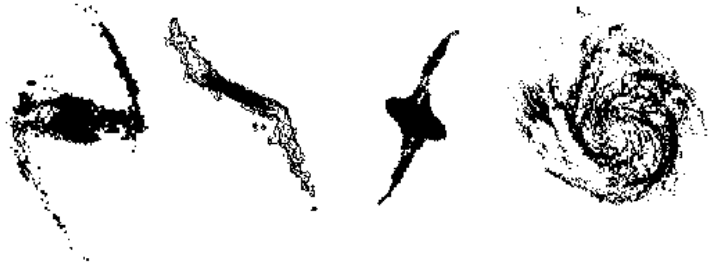}}
\caption{Dynamical processes constraining the external gas accretion,
illustrated by prototypical examples,
from left to right: frequency of bars in galaxies (NGC 1365), frequency
of warps (NGC 4013), polar rings (NGC 4650A), and lopsidedness (M101).}
\label{dyn-gas}
\end{figure}

Several dynamical processes can be naturally explained with 
external gas accretion, and could be used as constraints
on the amount of gas available, they are 
summarised in fig \ref{dyn-gas}.

\subsection{Bars and secular evolution}

From observations and numerical simulations in the last decades,
secular evolution has been shown to be driven by bars and spirals
(e.g. Kormendy \& Kennicutt 2004). In particular, the bar gravity
torques produce radial gas inflow, generating star formation.
Dynamical instabilities then regulate themselves, since the
gas inflow itself can destroy the bar. The bar is destroyed by two main
mechanisms: first the central mass concentration built after the
gas inflow, destroys the orbital structure sustaining the bar,
scatter particles and push them on chaotic orbits (Hasan et al 1990, 1993,
Hozumi \& Hernquist 1999). Second, the gas inflow itself weakens
the bar, since the gas loses its angular momentum to the stars
forming the bar (Bournaud \& Combes 2004). This increases
the angular momentum of the bar wave, in decreasing
the eccentricity of the orbits.
This bar destruction is reversible, since the central mass
concentration is then not strong enough to prevent 
bar formation.

Secular evolution then includes several bar episodes in a galaxy 
life-time. A spiral galaxy rich in gas (at least 5\% of the disk mass)
is unstable with respect to bar formation. Gravity torques are then
efficient to drive the matter inwards. The galaxy morphological type
evolves towards early-types, the mass is concentrated, the bulge
is developped, through horizontal and vertical resonances.
This weakens the bar, and when the galaxy becomes again 
axi-symmetric, gas can be accreted from the outer parts by
viscosity (Bournaud \& Combes 2002). The gas accretion,
if significant with respect to the disk mass, can reduce
the bulge-to-disk ratio (by replenishing the disk), and
make the galaxy disk unstable again to a new bar.

Several bars can successively produce secular evolution of
spiral galaxies, if there is enough external gas accretion.
This can be used to quantify the amount of accretion in 
a galaxy life-time.

To estimate the number of bars that has occured in
a typical galaxy, and the time spent in a bar phase, Block
et al (2002) have estimated quantitatively the frequency
of bars and  $m=2$ spiral components in a sample of 163
spiral galaxies, observed in the near-infrared (Eskridge et al
2002). In this band, it is more easy to identify the bars
in the old stellar component, with limited dust extinction.

The bar strength $Q_b$ has been estimated from the gravitational
potential, derived from the NIR images, assuming constant
mass-to-light ratios. The main surprising result 
in the histogram of bar strength is the hole at low
values (lack of axi-symmetric galaxies), and a long
tail at high values, corresponding
to a large number of strongly barred galaxies
(see also Whyte et al 2002, Buta et al 2004).

It can be concluded from the comparison of bar life-times
given by numerical simulations, and the observation of
bar frequency, that bars have to be reformed, and this
can only be through significant accretion of external gas
(see Bournaud \& Combes, this volume).

\subsection{Warps}
Most spiral galaxies reveal a warped disk, particularly spectacular in the HI component
 (Briggs 1990), although it is still frequent in the optical disk (Reshetnikov 
\& Combes 1998). Many physical mechanisms have been studied to explain the origin
 of these warps (see the review by  Binney 1992), but the high frequency of warps 
is only accounted for by external gas and dark matter accretion. Matter accreted
 in the outer parts of a dark matter halo, with an angular momentum misaligned with
 that of the inner parts, car re-oriented the whole system in 7-10 Gyr, and the 
corresponding disk will show an integral-sign warp during the process (Jiang \& 
Binney 1999). Intergalactic gas accretion is sufficient to produce a torque 
corresponding to the observed warps, with the infall amplitude already required 
by evolutionary chemical models (Lopez-Corredoira et al 2002). During some 
transient phase, a U-shape warp is predicted, combining m=0 and m=1 perturbations,
 and its frequency corresponds to the observations. The galaxy has just to move in 
an accretion flow of average density of 10$^{-4}$ cm$^{-3}$, and velocity 100km/s. 

Bullock et al (2001) have explored the angular momentum distribution in a large
number of dark matter haloes, from a cosmological simulation. They found an almost 
universal form, with a moderate misalignment in the outer parts, that could 
also favor warps in the baryonic matter. However, both the halo growth through 
merger of smaller units, and quiescent and continuous accretion are able to produce the
same universal profile, so this will not be discriminant.

\subsection{Polar rings}

Polar ring galaxies are particular objects, composed of a primary
galaxy in general of early-type (lenticular or elliptical), and a
polar ring or disk, where the matter is orbiting in a nearly
perpendicular plane. The polar material is quite gas-rich, but
also composed of stars, of age larger than a few Gyrs, implying that the
polar disk is stable. Indeed, in any of the various scenarios 
advanced to account for this structure, the bulk of the stars form later
from the gas settled in the polar plane. The probability for a 
galaxy to have a polar ring has been estimated to $\sim$ 5\%, given
the detection biases, that these systems are better detected
edge-on (Whitmore et al 1990).

The two main scenarios to form polar rings are mergers of two galaxies,
or external gas accretion, this gas coming either from a passing-by
companion, or from a cosmic gaseous filament. In the merging scenario
(Bekki 1997, 1998), a special geometry is required for the two
interacting system, it must be a head-on collision at low velocity,
with the two galaxy planes nearly perpendicular to each other.
In the accretion scenario, gas may be accreted from an unbound
companion passing by (Schweizer et al. 1983, Reshetnikov \& 
Sotnikova 1997), in a nearly polar orbit. The companion must be gas-rich,
and may also be more massive than the primary. 
But gas could also come from a gaseous cosmic filament, during the
long formation  period of a giant galaxy, since the various filaments
have not the same orientation. This is the extreme case of the
warp formation mechanism, invoked in the last subsection, where 
the various accreted gas have not aligned angular momentum.
Intermediate cases between warps and polar rings, where rings
of gas orbit at around 45$^\circ$, are observed (for instance NGC 660).
Polar rings around galactic-like objects are frequently seen in 
cosmological simulations (Semelin \& Combes 2004).

From a large number of numerical simulations, exploring the
various geometrical or mass parameters of the two scenarios,
it has been shown that the accretion scenario is more 
likely to produce polar rings, corresponding to observations,
than the merger scenario (Bournaud \& Combes 2003). In addition
to this statistical argument, some prototypical cases, like
NGC 4650A,  support the accretion hypothesis, since they do
not possess a halo of very old stars, expected in the merger
scenario.

\subsection{Lopsidedness}
The observation of extended disks of neutral hydrogen
(through the HI line at 21cm) around most spiral disks, has
revealed a large frequency of asymmetries and lopsidedness.
In their compiled sample of 1700 galaxies, Richter \& Sancisi (1994) 
found that at least 50\% of spiral galaxies are lopsided,
and have a characteristic signature in their global HI
spectrum. This percentage is even higher in late-type galaxies, where
Matthews et al (1998) found a frequency of 77\% of HI distorted
profiles. 
The asymmetries affect also the stellar disk, as shown by
Rix \& Zaritsky (1995) and Zaritsky \& Rix (1997) on 
near-infrared images of galaxies. About 20-30\% of
galaxies have a significantly perturbed stellar disk,
quantified by the $m=1$ Fourier term in their potential
or density distribution (Kornreich et al 1998, Combes 
et al 2004).

Since most of these galaxies are isolated without obvious
sources of perturbations (e.g. Wilcots \& Prescott 2004), 
the interpretation was searched
in a possible long life-time of these features.
Baldwin et al. (1980) proposed that the lopsidedness comes
from $m=1$ kinematic waves
build from off-center elliptical orbits that may persist a long time
against differential precession. 
Although this can prolonge the life-time significantly, 
since the winding out by differential precession is quite long
in the outer parts of galaxies, they
conclude that it is still not sufficient
to explain the high frequency of lopsidedness in neutral gas disks.

Alternatively, the perturbations could come from recent minor mergers, explaining
the non-correlation with the presence of companions (Zaritsky \& Rix 1997).
In this hypothesis,  lopsidedness can be used to constrain the 
merger frequency, which appears very high.
Indeed, Walker et al (1996) have estimated through N-body simulations
that the life-time of perturbations are of the order of the Gyr.
 A high merging frequency may enter in conflict with other
observations, for instance the presence of thin stellar disks in
spiral galaxies (Toth \& Ostriker 1992).

External gas accretion could solve the problem of the high frequency
of lopsidedness, since it is likely that in many cases, the 
accretion is asymmetric.

\section{Environmental effects}

The relative role of galaxy merging and external gas accretion must depend
strongly on environment. In particular, in rich clusters, mergers have
considerably increased the fraction of spheroids and ellipticals, and
tidal interactions and ram pressure have stripped and heated the cold
gas around galaxies, so that external accretion will be reduced or suppressed.
To quantify theses effects, we consider successively  the influence
of environment on star formation rate, morphological types, and
dynamical features, such as bars or warps.

\subsection{Star formation history}
 High resolution images with the HST, followed by spectroscopic
surveys in a dozen galaxy clusters at redshift around 0.3-1. have allowed to
follow galaxy evolution and in particular their star formation history
as a function of time, and also position in the cluster (Oemler et al 1997,
Dressler et al. 1999, Poggianti et al 1999).

Rich clusters of galaxies, and especially their dense cores,
are well known to be dominated by early-type galaxies without
star formation, or even passively evolving and anemic spirals,
conspicuous for their low star-formation rate. To find star formation
it is necessary to look back in time, at redshift at least larger
than 0.2, where there exists a larger fraction of blue galaxies
(B-O effect, Butcher \& Oemler 1978, 1984). 
The recent HST images have shown that these blue galaxies are
disk galaxies in majority perturbed by tidal interactions, strongly
suggesting that galaxy interactions are still triggering star formation
activity around $z=0.5$ (Lavery \& Henry 1988).

There is thus a strong evolution effect in clusters in a recent past.
Moreover, detailed spectroscopic studies have shown that $z \sim 0.5$ clusters
possess a large fraction of peculiar galaxies, likely post-starburst,
called E+A (or k+a), devoid of emission lines (and therefore with
no current star formation), but very strong Balmer absorption lines.
(Dressler et al 1999, Poggianti et al 1999). This means that they have 
a large fraction of A stars, implying that the galaxy was experiencing
a strong starburst that has just been abruptly interrupted.
These galaxies are in majority disk-dominated. Their fraction is
about 20\%, much larger than in the field, and in the clusters at $z=0$
(Dressler et al 1999). 

That star formation is quenched in clusters is revealed by the
low fraction ($\sim$ 10\% ) of H$\alpha$ emitters;
star formation and morphological evolution in cluster galaxies appear 
to be largely decoupled (Couch et al 2001). On the contrary,
the star formation is continuing in groups. About 55\% of galaxies have been
cataloged in groups at $z < 0.1$ in the 2dF and SDSS surveys,
and the  H$\alpha$ detection rate and equivalent width varies
strongly and continuously with galaxy density (Balogh et al 2004).

\begin{figure}[ht]
\centerline{\includegraphics[width=10cm]{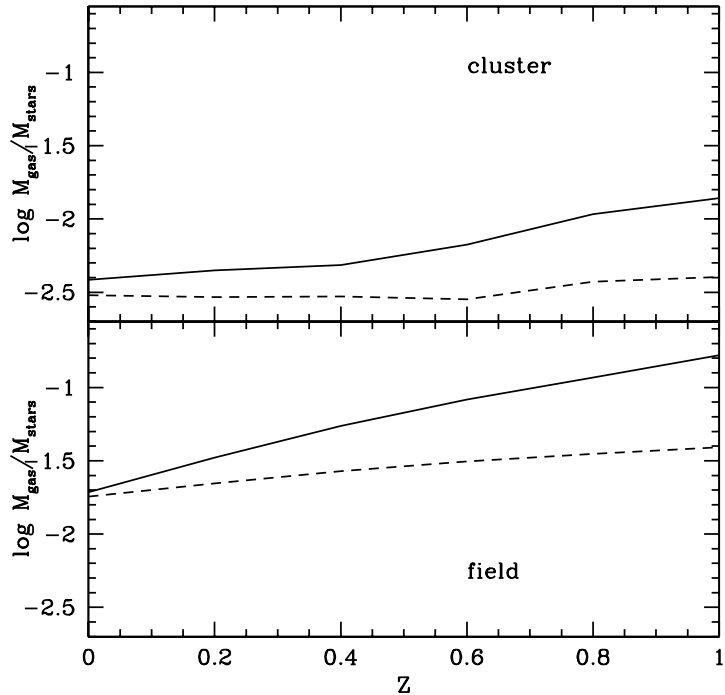}}
\caption{The ratio of gas mass to stellar mass as
a function of redshift for galaxies larger than 
3 10$^{10}$ M$_\odot$ in clusters (top) and in the field
(bottom), from semi-analytical models by Diaferio et al (2001).
The dashed lines correspond to SFR proportional to 
the cold gas mass over the dynamical time. The solid lines 
include in addition a dependency of the SFR with redshift
as (1+z)$^{-1.5}$.}
\label{diaf-f1}
\end{figure}

The star formation rate, as traced by the H$\alpha$ line, is strongly dependent
on the density of galaxies (Lewis et al 2002, 2dF survey). There is a tight correlation 
between SFR and local projected density, as soon as
the density is above $\Sigma$ = 1 galaxy/Mpc$^2$, independent of the size
of the structure.   In clusters, the field star formation
rate is only recovered at about 3 times the virial radius. 
G\'omez et al (2003) find also a strong SFR-$\Sigma$ relation
with the early data release of the SDSS, the SF-quenching effect
being even more noticeable for strongly star-forming galaxies.
The same break of the SFR-$\Sigma$ relation is observed
at 1 galaxy/Mpc$^2$. This relation is somewhat linked to the morphological
type-density (T-$\Sigma$) relation, but cannot be reduced to it, 

The spatial distribution of star forming galaxies is also quite clear
in clusters. There is a clear radial gradient, the star formation
being more active in the outer parts (Balogh et al 1999).  There
is a smooth transition from a blue, disk-dominated galaxy population in the
outskirts, to a red, bulge-dominated, galaxy population in the cluster
cores. Although this is influenced by the various transformation
processes occuring in clusters (tidal interaction, mergers, harrassment,
ram-pressure..), this has been mainly interpreted in terms of the building up
of the clusters themselves, i.e. gas-rich galaxies  continue to fall into
the cluster, and contribute to the star-formation activity there, but then
their gas replenishment is stopped when these galaxies arrive into the core
(Diaferio et al. 2001). In a model where the morphology of galaxies are
assumed to depend only on their merger history, where mergers lead to
spheroid and bulges, and subsequent gas cooling replenishes disks,
it is possible to reproduce the morphological gradient, and star-formation
gradient observed. Galaxies in clusters are much sooner devoid of their
gas, and their gas fueling is stopped, as shown in Figure \ref{diaf-f1}.

\subsection{Morphological types}
 There is a strong morphological segregation as a function
of projected density $\Sigma$ at $z=0$ (Dressler 1980), essentially
the fraction of spirals fall from two thirds to less than one
third to the benefit of lenticulars and ellipticals.
This segregation has also been observed at $z=0.4$ (or 5Gyrs
ago), where the fraction of ellipticals is the same
as at $z=0$ (Dressler at al 1997). The difference is
in the fraction of lenticulars, which is less than at $z=0$. 
This suggests that the formation of elliptical galaxies 
occurs before the formation 
of rich clusters, probably in the loose-group phase or earlier. 
Lenticulars are generated in large numbers only after cluster 
virialization, through the various galaxy transformation processes,
tidal interactions, mergers, ram-pressure stripping and sudden
halt in gas fueling (Poggianti et al 2001).
These results have been confirmed and precised with the local Sloan survey
by Goto et al (2003);  at least two mechanisms are required to explain
the morphological segregation: first at low density in the outskirts
of the cluster, the gas supply is stopped, and star formation halted.
Second at high density, in the cluster core, galaxy interactions
and merging must explain the high frequency of early-types.
The segregation is also found quite similar at $z=0.5$ and locally,
confirming that ellipticals in clusters have formed earlier. This corresponds to
numerical simulations, which show that the halos located inside clusters 
form earlier than isolated halos of the same mass 
(Gottloeber et al 2001).

\begin{figure}[ht]
\centerline{\includegraphics[width=10cm]{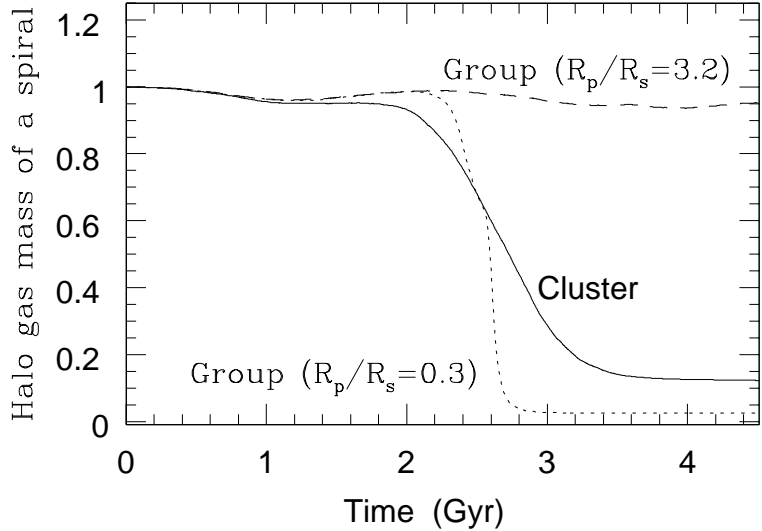}}
\caption{ Evolution with time of the halo gas mass
(normalized to the initial mass), for a spiral galaxy
orbiting a cluster with a pericenter distance $R_p$
equal to 3.2 the scale radius of the cluster $R_s$ =
230kpc (solid line).  By comparison, the dashed line
corresponds to  a comparable orbit in a group
with $R_s$ = 62kpc, and the dotted line, with
$R_p$ = 0.3 $R_s$ in this group (from Bekki et al 2002).}
\label{bek-f2}
\end{figure}

Many aspects of galaxy morphology in clusters can be
explained by evolutionary processes (see the reviews by
Moore 2003, Combes 2003), but the fact that
the gas supply is halted plays in all these processes a major role
(Shioya et al 2004 and fig \ref{bek-f2}): this gas shortening could happen
in two ways, either gradual from gas stripping, and star formation
rate dropping subsequently, or abruptly, through a starburst
triggered by an interaction, and lack of gas supply afterwards.
The second way is necessary to explain the high frequency
of post-starburst systems (k+a) in distant clusters, and
the low frequency of strong H$\alpha$ line emitters.

\subsection{Frequency of bars, warps}

Spiral galaxies are globally less abundant in
clusters, and are preferentially located in the outskirts. It has
long been known that clusters  are the site of a particular 
class of spirals, called "anemic", with low star formation,
smooth arms, and low gas content (van den Bergh 1976). 
In average, anemic galaxies have an HI deficiency of
a factor 4, but no CO emission deficiency, their
spiral structure is fuzzy and short-lived, since
new stars with low velocity dispersion are no longer formed out of the gas
(Elmegreen et al 2002).

The frequency of bars in cluster spirals has been studied in order
to test the influence of tidal interactions by
Andersen (1996) for the Virgo cluster. Barred galaxies appear
to be more centrally concentrated in the cluster core than
unbarred galaxies. This is inferred from both velocity
dispersions which are quite different. On the contrary, barred or unbarred 
lenticulars have the same radial distribution in the cluster.
Andersen (1996) suggests that tidal triggering by the cluster  itself
is the most
likely source of the enhanced fraction of barred spirals in the cluster center.
The same phenomenon is observed in Coma (Thompson 1981). 
Bars occur twice as often in the core region
of Coma as they do in the outer parts of the cluster.
This is certainly the consequence of tidal interactions
triggering bar instability (Gerin et al. 1990,
Miwa \& Noguchi 1998, Berentzen et al 2004).

At a global scale however, there does not appear to be
any significant difference between the frequency of bars in
clusters and in the field (van den Bergh, 2002).
Because bars are triggered by galaxy interactions
in the cluster cores, this implies that the suppression
of gas supply in clusters has a tendency to reduce bar
frequency to compensate.

It is difficult to observe warps around spiral galaxies in
clusters, because of their characteristic general HI deficiency.
About two thirds of galaxy clusters are HI deficient (Solanes et al. 2001).
Gas is particularly stripped in the outer parts of galaxies,
where warps develop. 

\subsection{Polar rings and asymmetries}

The environment of polar ring galaxies seems to be similar to that of normal galaxies
(Brocca et al 1997). There is no evidence of more companions for example. 
Some polar ring galaxies are observed in clusters (Taniguchi et al. 1986).
As for asymmetries, they are very frequent in clusters,  but this has to be
attributed to galaxy interactions;  in a Hubble time, each galaxy suffers
about 10 encounters with an impact parameter less than 10kpc,
due to the importance of substructures (Gnedin 2003).

\begin{figure}[ht]
\centerline{\includegraphics[width=10cm]{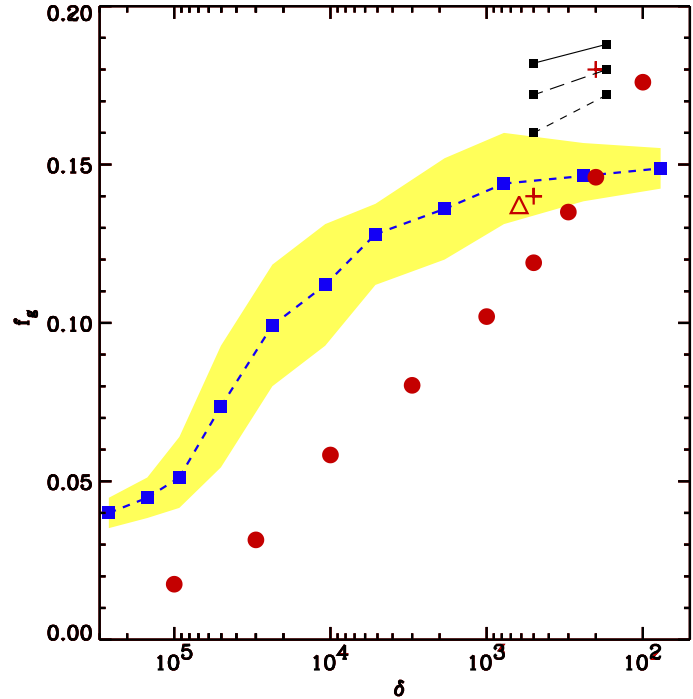}}
\caption{Distribution of the hot gas fraction f$_g$ in clusters of galaxies
as a function of $\delta$, the average density inside the radius r (normalised
to the critical density). The filled circles, empty triangle and crosses are the data,
while the squares connected by a dashed line are the theoretical predictions
(cosmological simulations from several groups),
with an assumed baryon fraction of f$_b = \Omega_b/\Omega_m$ =0.16;
other models (with or without stellar feedback)
are shown with f$_b = \Omega_b/\Omega_m$ =0.20 (small squares), 
from Sadat \& Blanchard (2001)}
\label{sadat01}
\end{figure}

\section{Summary}

The mass assembly of galaxies occurs through two main processes:
hierarchical merging of smaller entities, and more diffuse gas accretion.
The relative importance of the two processes cannot be 
easily found by cosmological simulations, since  
 many physical parameters such as gas dissipation, 
star formation and feedback,  are still unknown.
However, star forming histories in galaxies (age, kinematics
and metallicity), and also the dynamical states of galaxies
(bars, spirals, warps, polar rings, asymmetries..) can
constrain the role of the two processes.

Chemical evolution of normal spiral galaxies requires gas infall
of low-metallicity gas, in the proportion of a few solar masses per year.
The developpment of bars and spirals in galaxies constrain the 
amount of external gas accretion at about 10 M$_\odot$/yr. Bars are
transient features in galaxies, they exert gravity torques on the 
gas component, and drive mass to the center. A new bar can reform in 
disks that have been replenished in gas.  The large frequency of
gaseous warps in the outskirts of galaxies require also continuous
gas accretion, with misaligned angular momentum. When the accreted 
gas has perpendicular angular momentum, a polar ring can be formed.
Asymmetric gas accretion could be responsible for 
frequent lopsided galaxies.

In rich environments, the evolution is occuring at a much quicker
pace. Galaxy interactions and mergers are much more frequent, and
happen earlier in the Hubble time. The star formation activity
triggered by the mergers must have occurred early in dense groups
that coalesce then in galaxy clusters. Most of the star formation
is then quenched after redshift $z \sim$ 1. The only activity remaining
comes from the infall of field spiral galaxies in the outer parts
of the clusters. Secular evolution through external gas
accretion is slowed down or halted, since galaxies are stripped
from their gas, at large galactocentric radii. Most of the gas has
been shock-heated  at the virial temperature of the clusters,
and is observed in X-rays.  The mass fraction in hot gas reaches
the baryonic fraction in the outer parts (e.g. Sadat \& Blanchard 2001,
fig \ref{sadat01}, Valageas et al. 2002). Contrary to the field, it appears that 
hierarchical merging has dominated secular evolution in rich
galaxy clusters.

\begin{acknowledgments}
I wish to thank David Block for the organisation of this exciting conference.
\end{acknowledgments}

\end{document}